\documentclass[tradiabstract]{aa}

\newcommand{\bea}{\begin{eqnarray}}
\newcommand{\eea}{\end{eqnarray}}
\newcommand{\be}{\begin{equation}}
\newcommand{\ee}{\end{equation}}
\newcommand{\bc}{\begin{center}}
\newcommand{\ec}{\end{center}}
\newcommand{\ben}{\begin{enumerate}}
\newcommand{\een}{\end{enumerate}}
\newcommand{\bd}{\begin{description}}
\newcommand{\ed}{\end{description}}
\newcommand{\bmi}[1]{\begin{minipage}{#1 cm}}
\newcommand{\emi}{\end{minipage}}
\newcommand{\bmif}[1]{\begin{minipage}{#1\textwidth}}

\def\llabel#1{\label{sc:#1}}
\def\elabel#1{\label{eq:#1}}

\def\eck#1{\left\lbrack #1 \right\rbrack}

\def\rund#1{\left( #1 \right)}

\def\d{{\rm d}}

\def\eps{{\epsilon}}

\def\Real{{\rm I\mathchoice{\kern-0.70mm}{\kern-0.70mm}{\kern-0.65mm}%
  {\kern-0.50mm}R}}
\def\C{\rm C\kern-.42em\vrule width.03em height.58em depth-.02em
       \kern.4em}

\def\bx#1{\leavevmode\thinspace\hbox{\vrule\vtop{\vbox{\hrule\kern1pt
        \hbox{\vphantom{\tt/}\thinspace{\bf#1}\thinspace}}
      \kern1pt\hrule}\vrule}\thinspace}

\def\vc#1{{\mbox{\boldmath$#1$\unboldmath}}}
{\catcode`\@=11
\gdef\SchlangeUnter#1#2{\lower2pt\vbox{\baselineskip 0pt \lineskip0pt
  \ialign{$\m@th#1\hfil##\hfil$\crcr#2\crcr\sim\crcr}}}
}

\def\ueber#1#2{{\setbox0=\hbox{$#1$}%
  \setbox1=\hbox to\wd0{\hss$\scriptscriptstyle #2$\hss}%
  \offinterlineskip
  \vbox{\box1\kern0.4mm\box0}}{}}

\def\bx#1{\leavevmode\thinspace\hbox{\vrule\vtop{\vbox{\hrule\kern1pt
        \hbox{\vphantom{\tt/}\thinspace{\bf#1}\thinspace}}
      \kern1pt\hrule}\vrule}\thinspace}

{\catcode`\@=11
\gdef\SchlangeUnter#1#2{\lower2pt\vbox{\baselineskip 0pt \lineskip0pt
  \ialign{$\m@th#1\hfil##\hfil$\crcr#2\crcr\sim\crcr}}}
}

\def\ts{\thinspace}

\usepackage{graphicx}
\usepackage{txfonts}
\usepackage{natbib}
\usepackage{subfigure}
\usepackage{enumerate}
\begin{document}

   \title{Generalized shear-ratio tests: A new relation between
     cosmological distances, and a diagnostic for a redshift-dependent
     multiplicative bias 
     in shear measurements} 

   \author{Peter Schneider \inst{1} 
          }

   \institute{Argelander-Institut f\"ur Astronomie, Universit\"at
     Bonn, Auf dem H\"ugel 71, D-53121 Bonn, Germany\\
    peter@astro.uni-bonn.de}


 
  \abstract
  { We derive a new relation between cosological distances,  valid in
    any (statistically) isotropic space-time and independent of
    cosmological parameters or even the validity of the field equation
    of General Relativity. In particular, this relation yields an
    equation between those distance ratios which are the geometrical
    factors determining the strength of the gravitational lensing
    effect of mass concentrations. Considering a combination of weak lensing
    shear ratios, based on lenses at two different redshifts, and
    sources at three different redshifts, we derive a relation between
    shear-ratio tests which must be identically satisfied. A redshift-dependent
    multiplicative bias in shear estimates will violate this relation,
    and thus can be probed by this generalized shear-ratio
    test. Combining the lensing effect for lenses at three different
    redshifts and three different source redshifts, a relation between
    shear ratios is derived which must be valid independent of a
    multiplicative bias. We propose these generalized shear-ratio
    tests as a diagnostic for the presence of systematics in upcoming
    weak lensing surveys. } 

  \titlerunning{Generalized shear-ratio tests}

   \maketitle
%

\section{\llabel{Sc1}Introduction}
Weak gravitational lensing based on shear measurements are challenged
by obtaining an unbiased estimate of the shear from the observed
images of faint distant galaxies \citep[see,
e.g.,][]{STEP1,STEP2,GREAT3}. Biases in shear estimates are commonly
parametrized through
\be
\hat\gamma=(1+m)\gamma + c\;,
\elabel{1a}
\ee
where $\hat\gamma$ is the expectation value of the shear estimate,
$\gamma$ is the true shear -- a two-component quantity, conveniently
written as a complex number -- and $m$ and $c$ are the multiplicative
and additive biases, respectively. The additive bias itself is a
complex quantity and thus has a phase (or an orientation). As such, it
defines a direction on the sky. The presence of an additive bias can
thus be detected by correlating the estimated shear with other
quantities that define directions, e.g., the phase of the point-spread
function (PSF) for which any shape measurement must be corrected, signaling
an incomplete PSF-correction, or the
pixel grid of the detector. For the rest of the letter, we will ignore
the additive bias parameter.

In contrast to $c$, a multiplicative bias cannot be easily identified
from the data themselves. Such a bias is expected to arise from the
smearing correction of the PSF, pixelization, pixel noise
\citep[``noise bias''; e.g.,][]{2012MNRAS.424.2757M,
  2012AA...547A..98B}, and, depending on the method used for shear
estimates, insufficient knowledge of the distribution of intrinsic
brightness profiles of sources \citep[``underfitting bias'', or
``model bias, 
e.g.,][and references therein]{2010MNRAS.404..458V, 2010MNRAS.406.2793B,
  2013MNRAS.429.2858M, 2014MNRAS.438.1880B, 2015ApJ...807...87S}.

Thus, an absolute determination of $m$ from the data appears
extremely challenging. Possible methods for this include the use of
magnification information \citep{2010arXiv1009.5735R}, or the 
calibration of shear as measured from faint galaxy images with the
shear obtained from the lensed cosmic microwave background
\citep{2013arXiv1311.2338D}. If the multiplicative bias depends on
galaxy properties, such as color or size, a relative bias may be
detected in the data by splitting up the galaxy sample and comparing
the results.

In this letter, we propose a new method for detecting a
redshift-dependent multiplicative bias, which we call `generalized
shear-ratio test' (GSRT). It is a purely geometrical method, based on
a relation between distances in (statistically) isotropic universes
which we derive in Sect.\ts\ref{sc:Sc3}. In contrast to the `classical'
shear-ratio test (see Sect.\ts\ref{sc:Sc2}), which yields a geometrical
probe of the cosmological model assuming no multiplicative bias, the GSRT
is independent of the cosmological model, and even independent of the
validity of Einstein's field equation, but capable of detecting the
redshift-dependent bias $m$, as will be described in
Sect.\ts\ref{sc:Sc4}. Furthermore, we will derive a relation between
shear ratios which is even independent of a multiplicative bias, but
only depends on the redshifts of lens and source populations. As such,
this relation offers the opportunity to study the accuracy of
photometric redshifts in cosmological weak lensing surveys.

\section{\llabel{Sc2}The classical shear-ratio test}
Any observable weak lensing quantity that is linear in the shear
depends linearly on the surface mass density $\Sigma$ of the lens
\citep[see, e.g.,][]{bs-review, Kilbinger-review}. 
For a lens (or a population of lenses) at redshift $z_1$ and a source
population at redshift $z_2$, such a shear qunatity $S$ can be
expressed as 
\be
S=G {D(z_1,z_2)\over D(z_2)} \equiv G\,\beta(z_1,z_2)\;,
\elabel{1}
\ee
where $D(z_1,z_2)$ is the angular-diameter distance of the sources at
$z_2$ as seen from an observer at $z_1$, and $D(z_2)\equiv D(0,z_2)$
is the angular-diameter distance of the sources from us. The lensing
quantity $G$ is linear in the scaled dimensionless surface mass density 
\be
\kappa_{\rm sc}={4\pi G\over c^2}\,D(z_1)\,\Sigma\;,
\elabel{2}
\ee
and depends solely on the properties and distance of the lens
(population), but is independent of the source redshift. In
Eq.\ts (\ref{eq:1}), we have defined the distance ratio $\beta$, which for a
given lens characterizes the lensing strength as a function of
source redshift. 

The true image ellipticity $\eps$ of a background galaxy is an
unbiased estimate of the reduced shear $\gamma/(1-\kappa)$
\citep{SeitzS97}, where 
$\kappa=\kappa_{\rm sc}\,\beta(z_1,z_2)$ is the convergence of the lens
at $z_1$ for a source population at redshift $z_2$. Assuming that the
convergence is much smaller than 
unity, we neglect the difference between shear and reduced
shear. Then, an estimate of $S$ is obtained as a linear combination of
image ellipticities $\eps$, whose expectation value is 
\be
\Gamma(z_1,z_2) = G(z_1)\,\beta(z_1,z_2)\,M(z_2)\;,
\elabel{3}
\ee
where $M=(1+m)$, with $m$ being the aforementioned multiplicative bias
in shear measurements. For example, $\Gamma(z_1,z_2)$ could be a
weighted integral over separation of the tangential shear in
galaxy-galaxy lensing for lenses at $z_1$, measured for sources at
$z_2$, where the weight can be chosen such as to optimize the
signal-to-noise ratio of the measurement \citep[e.g.,][]{bs-review}.

Hence, the observable signal $\Gamma$ depends (i) on the properties of
the lens described by $G$, (ii) on the cosmological parameters through
the distance 
ratio $\beta$, and (iii) on the multiplicative bias in shear
measurements. Furthermore, it also depends (iv) on the reliability of
the estimates of (photometric) redshifts.

The classical shear ratio
test \citep{2003PhRvL..91n1302J, 2007MNRAS.374.1377T, 2015arXiv151203627K}
considers a lens (population) at redshift $z_1$, and two source
populations at redshifts $z_2$ and $z_3$. The ratio of their lensing
signals,
\be
R(z_1;z_2,z_3)\equiv {\Gamma(z_1,z_3)\over \Gamma(z_1,z_2)}
={\beta(z_1,z_3)\over \beta(z_1,z_2)}\,{M(z_3)\over M(z_2)}
\equiv B(z_1;z_2,z_3)\,{M(z_3)\over M(z_2)}\;,
\elabel{4}
\ee
eliminates the dependence on the lens properties. Due to its
dependence on the distance ratio $B$, the shear-ratio $R$ was
proposed as a probe for cosmological parameters. However, the
sensitivity of $B$ on, e.g., the equation-of-state parameter $w$ of
dark energy turns out to be rather weak, so that highly accurate
measurements of $R$ are needed to turn this into a competitive
cosmological probe. Furthermore, unbiased estimates of shear and
redshifts which enter the shear ratio test provide a huge challenge
for precision weak lensing experiments. It is therefore of
great interest to find observational probes for a potential
multiplicative bias in shear measurements and a bias of photometric
redshifts which is insensitive to assumptions about the cosmological
parameters. 

In the next section, we will derive relations between cosmological
distances which are independent of cosmological parameters and which
allow us to construct combinations of $R$ which carry no longer a
dependence on cosmology.

\section{\llabel{Sc3}Distance relation in isotropic universes}
The two-dimensional separation vector $\vc \xi$ between two
infinitesimally close light rays follows the geodesic deviation
equation 
\be
{\d^2 \vc\xi\over \d\lambda^2}={\cal T}(\lambda)\,\vc\xi(\lambda)\;,
\elabel{5}
\ee
where $\lambda$ is the affine parameter along the light rays, and
${\cal T}(\lambda)$ is the optical tidal matrix, which is determined
by the Ricci and Weyl tensors of the spacetime metric \citep[see,
e.g., 
Chap.\ts 3 of Schneider et al.\ \citeyear{SEF}, and][]{SSE}. In an
isotropic universe, the tidal part of ${\cal T}$ vanishes, so that
${\cal T}(\lambda)=T(\lambda)\,{\cal I}$, where ${\cal I}$ is the
two-dimensional unit matrix and $T(\lambda)$ is a scalar function,
proportional to the local density.

If the two light rays intersect at a vertex at $\lambda_i$ under an
angle $\vc\theta$, we can write
$\vc\xi(\lambda)=D_i(\lambda)\vc\theta$, where $D_i$ satisfies 
the differential equation
\be
{\d^2 D\over \d\lambda^2}=T(\lambda)\,D\;,
\elabel{6}
\ee
with initial conditions $D_i(\lambda_i)=0$, $(\d D_i/\d
\lambda)(\lambda_i)= \dot D_i$, where the latter value depends on the
choice of the affine parameter. By definition, $D_i(\lambda)$ is the
angular-diameter distance of a source at $\lambda$, seen from an
observer at $\lambda_i<\lambda$. 

The second-order differential equation (\ref{eq:6}) has two
independent solutions; we choose them to be $D_i(\lambda)$ and
$D_j(\lambda)$, i.e., the angular-diameter distances as measured from
observers located at $\lambda_i$ and 
$\lambda_j\ne \lambda_i$, respectively.
A third solution, $D_k(\lambda)$, must necessarily be a
linear combination of the other two. If we write it in the form
\[
D_k(\lambda)=C\eck{D_j(\lambda)D_i(\lambda_k)-D_j(\lambda_k)D_i(\lambda)}\;,
\]
where $C$ is a constant,
then we see that the first initial condition is satisfied,
$D_k(\lambda_k)=0$. For the second initial condition, we have
\bea
{\d D_k\over \d\lambda}(\lambda_k)&=&\dot D_k
=C \eck{{\d
    D_j\over\d\lambda}(\lambda_k)D_i(\lambda_k)-D_j(\lambda_k){\d
    D_i\over \d\lambda}(\lambda_k)} \nonumber \\
&=& C W(\lambda_k)\;,
\elabel{7}
\eea
where $W(\lambda)$ is the Wronskian
\[
W(\lambda)={\d D_j\over\d\lambda}\,D_i - D_j\,{\d D_i\over\d\lambda}\;.
\]
From Eq.\ts (\ref{eq:6}), one sees directly that $\d W/\d\lambda=0$, i.e.,
$W(\lambda)$ is constant. We can calculate this constant considering
$W$ at $\lambda_j$, which yields 
\[
W=W(\lambda_j)=D_i(\lambda_j)\,\dot D_j\;,
\]
yielding $C=(\dot D_k / \dot D_j)\,
  D_i^{-1}(\lambda_j)$. Therefore, 
\be
D_k(\lambda)= {\dot D_k \over \dot D_j}\,{1\over D_i(\lambda_j)}
\eck{D_j(\lambda)D_i(\lambda_k)-D_j(\lambda_k)D_i(\lambda)}\;.
\elabel{8}
\ee
Thus, we have expressed the angular-diameter distance as seen from an
observer at affine parameter $\lambda_k$ by the corresponding
expression for the angular-diameter distance for observers at
$\lambda_i$ and $\lambda_j$. The resulting expression contains the
initial conditions at $\lambda_j$ and $\lambda_k$. If we now
specialize Eq.\ts (\ref{eq:8}) to the case of $\lambda_i=0$, with the
corresponding angular-diameter distance denoted
as $D(\lambda)$, we obtain
\be
D_k(\lambda)= {\dot D_k \over \dot D_j}\,{1\over D(\lambda_j)}
\eck{D_j(\lambda)D(\lambda_k)-D_j(\lambda_k)D(\lambda)}\;.
\elabel{9}
\ee
Combining Eqs.\ts (\ref{eq:8}) and (\ref{eq:9}), we then find
\bea
{\dot D_j \over \dot D_k}\,D_k(\lambda) 
&=&
{1\over D_i(\lambda_j)}
\eck{D_j(\lambda)D_i(\lambda_k)-D_j(\lambda_k)D_i(\lambda)}\nonumber \\
&=&{1\over D(\lambda_j)}
\eck{D_j(\lambda)D(\lambda_k)-D_j(\lambda_k)D(\lambda)}\;.
\elabel{10}
\eea
The final equality does not contain the initial conditions anymore. It
is valid for any spacetime which is isotropic around the observer
through which the radial light bundle passes. Furthermore, this
equation is not referring to the Einstein field equation, and is
therefore independent of the validity of General Relativity, as long a
spacetime is characterized by a metric.

If, in addition to isotropy, we assume a spatially homogeneous
spacetime, which can then be expressed in the form of the
Robertson--Walker metric, the initial conditions are $\dot
D_i=(1+z_i)$, where $z_i$ is the redshift with which sources at
$\lambda_i$ are seen by the observer at $\lambda=0$ (or redshift
zero). For this, we have chosen a parametrization of the affine
parameter such that it locally coincides with the comoving distance as
seen from the observer. In what follows, we assume that the isotropic
universe is such that an invertible relation between affine parameter
and redshift exists, so that redshift can be used to label the
relative order of objects (in the sense that an object at $z_2$ lies
behind one at $z_1<z_2$). This excludes, for example, bouncing models,
for which $z(\lambda)$ is not a monotonic function.

We will now identify $D_i(\lambda_j)$ as the distance $D(z_i,z_j)$,
and $D(\lambda_j)$ as $D(z_j)$, i.e., the quantities occurring in the
distance ratio $\beta$ (see Eq.\ts\ref{eq:1}). 
Setting $\lambda=\lambda_l$ in Eq.\ts (\ref{eq:10}) and dividing the
resulting expression by $D_k D_l$, we obtain with the definition
(\ref{eq:1}) that
\be
\beta_{jl}\beta_{ik}-\beta_{jk}\beta_{il}
=\beta_{ij}\rund{\beta_{jl}-\beta_{jk}} \;,
\elabel{11}
\ee
an expression between distance ratios $\beta_{ij}\equiv\beta(z_i,z_j)$.
This can be manipulated, by division through $\beta_{ij}\beta_{jk}$,
to contain just ratios of $\beta$'s with the first index being the
same -- or if we set $0\le z_i < z_j<z_k,z_l$, with the lower redshift
being the same. Hence, the resulting expression only contains the
ratios $B$ that were defined in Eq.\ts (\ref{eq:4}),
\be
B(z_j;z_k,z_l)={B(z_i;z_j,z_l)-1 \over B(z_i;z_j,z_k)-1}\;.
\elabel{12}
\ee
Of course, the relation (\ref{eq:11}) or (\ref{eq:12}) must hold in a
Robertson--Walker metric. The proof of this is outlined in the
appendix.

\section{\llabel{Sc4}Generalized shear ratio tests}
The relation between distances, as derived in the previous section, can
now be used to obtain ratios between shear observables
$\Gamma$. According to Eq.\ts (\ref{eq:4}), shear ratios depend on the ratios
of multiplicative bias factors, hence they can not be used to
determine the absolute multiplicative bias, but only their relative
values. 

Consider now two populations of lenses at redshifts $z_1$ and $z_2$,
and sources at redshifts $z_2$, $z_3$ and $z_4$.  Combining Eq.\ts
(\ref{eq:4}) with Eq.\ts (\ref{eq:12}), setting $(i,j,k,l)=(1,2,3,4)$,
we find that
\be
R(z_2;z_3,z_4)={R(z_1;z_2,z_4)-M(z_4)/M(z_2) 
\over R(z_1;z_2,z_3)-M(z_3)/M(z_2) } \;.
\elabel{13}
\ee
This relation between ratios of shear observables is independent of
the cosmological model and the validity of the field equation of
General Relativity, and contains only ratios of the multiplicative
bias at the three source redshifts. Hence we have found a
cosmology-independent shear ratio test. The special form of
Eq.\ts (\ref{eq:13}) has been chosen by assuming that the sources at the
nearest redshift (here $z_2$) can be measured most reliably, since
these sources probably will on average be larger than those at higher
redshift. Thus, we normalize the multiplicative bias by that for
sources at $z_2$. 

The redshift dependence may not be the primary dependence of the
multiplicative bias; rather, one might suspect that it depends on
image size, signal-to-noise ratio, galaxy color etc. Those
dependencies can be studied by splitting the source sample. A test
based on Eq.\ts(\ref{eq:13}) may serve as an additional or
complemetary `sanity check'.

Furthermore, we can consider three lens populations at redshift $z_1$,
$z_2$, and $z_3$, and three source populations at $z_3$, $z_4$, and
$z_5$. Setting $(i,j,k,l) \to (1,3,4,5)$ in (\ref{eq:13}), we find with
(\ref{eq:4}) that
\be
R(z_3;z_4,z_5)\eck{R(z_1;z_3,z_4)-{M(z_4)\over M(z_3)}}
=R(z_1;z_3,z_5)-{M(z_5)\over M(z_3)} \;.
\elabel{14}
\ee
On the other hand, by choosing $(i,j,k,l) \to (2,3,4,5)$, we get
\be
R(z_3;z_4,z_5)\eck{R(z_2;z_3,z_4)-{M(z_4)\over M(z_3)}}
=R(z_2;z_3,z_5)-{M(z_5)\over M(z_3)} \;.
\elabel{15}
\ee
Subtracting (\ref{eq:15}) from (\ref{eq:14}), the ratios of the
multiplicative bias factors cancel out, and we are left with
\bea
R(z_3;z_4,z_5)&&\!\!\!\!\!\!\!\!\!\!
\eck{R(z_2;z_3,z_4)-R(z_1;z_3,z_4)} \nonumber \\
=&&\!\!\!\!\!\!\!\!\!\!R(z_2;z_3,z_5)-R(z_1;z_3,z_5) \;,
\elabel{16}
\eea
a relation that just contains the observable shear ratios! 
Hence, this relation between observables must be obeyed; its violation
would indicate a problem with the redshift estimates of the sources
and/or lenses. Whereas (\ref{eq:13}) probes a combination of the
multiplication bias and redshift estimates,
(\ref{eq:16}) solely probes the latter; in this way, these two effects
can be disentangled.

Curiously, these relations are insensitive to getting the redshifts
right. For example, Eq.\ts (\ref{eq:13}) is valid for all $z_i$, $1\le
i\le 4$. If one believes that the sources have a redshift $z_4$, but
in reality are located at redshift $z_4'\ne z_4$, the relation
(\ref{eq:13}) still remains true. What is important, though, is that
the lens redshift $z_2$, occurring on the left-hand side, is the same
as the source redshift $z_2$ appearing on the right. We will study in
a future work how errors in the mean, the dispersion about the mean,
and catastrophic outliers in photometric redshift estimates affect
these GSRTs.

\section{\llabel{Sc5}Discussion}
We have derived two relations between measured shear ratios which are
independent of the cosmological model, following from two newly
derived relations between distances in isotropic space-times.
The first of these GSRTs,
Eq.\ts(\ref{eq:13}), relates shear ratios to ratios of the
multiplicative bias. Hence, if 
\[
{\cal R}_{1234}\equiv R(z_2;z_3,z_4)\eck{R(z_1;z_2,z_3)-1}-R(z_1;z_2,z_4)+1
\]
deviates from zero, a redshift-dependent multiplicative bias is a
likely origin. Another possible reason for ${\cal R}_{1234}\ne 0$
could be found in problems with redshift estimates of the
source galaxies. The latter is probed by the relation (\ref{eq:16}),
which is independent of the multiplicative bias.

In order to turn the GSRT into a diagnostic tool for future weak
lensing applications, several issues need to be studied before. First,
the sensitivity for measuring
statistically significant deviations of ${\cal R}_{1234}$ from zero
depends on the number of lenses and sources available for this test,
i.e., on the sky area of the weak lensing survey and its depth. It may
be conceivable that the lens population is selected based on
spectroscopic redshifts. For example, the SDSS has obtained far more
than one million spectroscopic galaxy redshifts
\citep{2015ApJS..219...12A}, and the upcoming experiments 4MOST
\citep{2014SPIE.9147E..0MD} and DESI \citep{2014SPIE.9147E..0SF} will
increase this number by more than an order of
magnitude. Alternatively, or in combination, the lens population can
be chosen as galaxy clusters, where the upcoming eROSITA X-ray survey
mission is expected to detect some $10^5$ clusters
\citep[e.g.,][]{2012MNRAS.422...44P}. 

Second, a practical application of the GSRT requires binning of the lens
and source samples into redshift bins. Optimal strategies for this
binning need to be developed. The estimators will also be affected by
the dispersion of the photometric redshift estimates which need to be
accounted for. 

Third, we have assumed an isotropic universe. Our universe is clearly
not isotropic on small scales, and hence the Weyl part of the optical
tidal matrix $\cal T$ in (\ref{eq:5}) does not vanish. However, the
difference between the 
mean distances in a locally inhomogeneous universe and the distances
in the corresponding smooth universe is extremely small \citep[see
  Sect.\ts 4.5 of Schneider et al.\ \citeyear{SEF}, and references
  therein, as well as the recent detailed discussion
  by][]{2016MNRAS.455.4518K}. 

\begin{acknowledgements}
  I would like to thank Stefan Hilbert, Hendrik Hildebrandt, Jeffrey
  Kotula, Patrick Simon and Sandra Unruh for 
  discussions and/or comments on the manuscript. 
  This work was supported by the Transregional Collaborative
  Research Center TR33 `The Dark Universe' of the German
  \emph{Deut\-sche For\-schungs\-ge\-mein\-schaft, DFG\/}.
\end{acknowledgements}

\begin{appendix}
\section{The validity of Eq.\ts(\ref{eq:11}) in a Robertson--Walker
  metric} 

In this appendix we outline the proof that Eq.\ts(\ref{eq:11}) holds
in a Robertson--Walker metric.
This can be shown using the following steps: (i) For the
distance ratios $\beta$, one can either use the angular-diameter
distances, or the comoving angular-diameter distances. Hence,
\[
\beta(z_i,z_j)={{\rm sn}\rund{\sqrt{|K|}\,[\chi_j-\chi_i]}\over
{\rm sn}\rund{\sqrt{|K|}\, \chi_j}}
={{\rm sn}(x_j-x_i)\over{\rm sn}(x_j)} \;,
\]
where the function ${\rm sn}(x)$ is either the sine or the hyperbolic
sine, depending on the sign of the curvature parameter $K$, and in the
limiting case of $K=0$, it is the identity. Furthermore, $\chi_j$ is
the comoving distance to redshift $x_j$, and we defined
$x_j=\sqrt{|K|}\chi_j$. (ii) Using the above expression, we can write
Eq.\ts(\ref{eq:11}) after multiplication by ${\rm sn}(x_k)\,{\rm
  sn}(x_l)$ as
\bea
{\rm sn}(x_l-x_j)\!\!\!\!\!\!&&\!\!\!\!\!\!\!{\rm sn}(x_k-x_i)
- {\rm sn}(x_k-x_j)\;{\rm sn}(x_l-x_i)\nonumber \\
&=&\!\!\!\!\!{{\rm sn}(x_j-x_i) \over {\rm sn}(x_j)}
\eck{{\rm sn}(x_l-x_j)\,{\rm sn}(x_k) 
- {\rm sn}(x_k-x_j)\,{\rm sn}(x_l) }  \;. \nonumber
\eea
(iii) Next we make use of the addition theorem, ${\rm sn}(a-b)={\rm
  sn}(a)\,{\rm cn}(b)-{\rm cn}(a)\,{\rm sn}(b)$, where ${\rm cn}(x)$ is
either $\cos(x)$, $\cosh(x)$, or $\equiv 1$, depending on $K$. After
applying the addition theorem to the foregoing equation, a
term-by-term comparison of both sides shows its validity. 

\end{appendix}

\bibliographystyle{aa}
\bibliography{MSDbib}

\begin{thebibliography}{25}
\expandafter\ifx\csname natexlab\endcsname\relax\def\natexlab#1{#1}\fi

\bibitem[{{Alam} {et~al.}(2015){Alam}, {Albareti}, {Allende Prieto}, {Anders},
  {Anderson}, {Anderton}, {Andrews}, {Armengaud}, {Aubourg}, {Bailey}, \&
  et~al.}]{2015ApJS..219...12A}
{Alam}, S., {Albareti}, F.~D., {Allende Prieto}, C., {et~al.} 2015, \apjs, 219,
  12

\bibitem[{{Bartelmann} \& {Schneider}(2001)}]{bs-review}
{Bartelmann}, M. \& {Schneider}, P. 2001, \physrep, 340, 291

\bibitem[{{Bartelmann} {et~al.}(2012){Bartelmann}, {Viola}, {Melchior}, \&
  {Sch{\"a}fer}}]{2012AA...547A..98B}
{Bartelmann}, M., {Viola}, M., {Melchior}, P., \& {Sch{\"a}fer}, B.~M. 2012,
  \aap, 547, A98

\bibitem[{{Bernstein}(2010)}]{2010MNRAS.406.2793B}
{Bernstein}, G.~M. 2010, \mnras, 406, 2793

\bibitem[{{Bernstein} \& {Armstrong}(2014)}]{2014MNRAS.438.1880B}
{Bernstein}, G.~M. \& {Armstrong}, R. 2014, \mnras, 438, 1880

\bibitem[{{Das} {et~al.}(2013){Das}, {Errard}, \&
  {Spergel}}]{2013arXiv1311.2338D}
{Das}, S., {Errard}, J., \& {Spergel}, D. 2013, ArXiv:1311.2338

\bibitem[{{de Jong} {et~al.}(2014){de Jong}, {Barden}, {Bellido-Tirado},
  {Brynnel}, {Chiappini}, {Depagne}, {Haynes}, {Johl}, {Phillips}, {Schnurr},
  {Schwope}, {Walcher}, {Bauer}, {Cescutti}, {Cioni}, {Dionies}, {Enke},
  {Haynes}, {Kelz}, {Kitaura}, {Lamer}, {Minchev}, {M{\"u}ller}, {Nuza},
  {Olaya}, {Piffl}, {Popow}, {Saviauk}, {Steinmetz}, {Ural}, {Valentini},
  {Winkler}, {Wisotzki}, {Ansorge}, {Banerji}, {Gonzalez Solares}, {Irwin},
  {Kennicutt}, {King}, {McMahon}, {Koposov}, {Parry}, {Sun}, {Walton},
  {Finger}, {Iwert}, {Krumpe}, {Lizon}, {Mainieri}, {Amans}, {Bonifacio},
  {Cohen}, {Fran{\c c}ois}, {Jagourel}, {Mignot}, {Royer}, {Sartoretti},
  {Bender}, {Hess}, {Lang-Bardl}, {Muschielok}, {Schlichter}, {B{\"o}hringer},
  {Boller}, {Bongiorno}, {Brusa}, {Dwelly}, {Merloni}, {Nandra}, {Salvato},
  {Pragt}, {Navarro}, {Gerlofsma}, {Roelfsema}, {Dalton}, {Middleton}, {Tosh},
  {Boeche}, {Caffau}, {Christlieb}, {Grebel}, {Hansen}, {Koch}, {Ludwig},
  {Mandel}, {Quirrenbach}, {Sbordone}, {Seifert}, {Thimm}, {Helmi}, {trager},
  {Bensby}, {Feltzing}, {Ruchti}, {Edvardsson}, {Korn}, {Lind}, {Boland},
  {Colless}, {Frost}, {Gilbert}, {Gillingham}, {Lawrence}, {Legg}, {Saunders},
  {Sheinis}, {Driver}, {Robotham}, {Bacon}, {Caillier}, {Kosmalski}, {Laurent},
  \& {Richard}}]{2014SPIE.9147E..0MD}
{de Jong}, R.~S., {Barden}, S., {Bellido-Tirado}, O., {et~al.} 2014, in
  \procspie, Vol. 9147, Ground-based and Airborne Instrumentation for Astronomy
  V, 91470M

\bibitem[{{Flaugher} \& {Bebek}(2014)}]{2014SPIE.9147E..0SF}
{Flaugher}, B. \& {Bebek}, C. 2014, in \procspie, Vol. 9147, Ground-based and
  Airborne Instrumentation for Astronomy V, 91470S

\bibitem[{{Heymans} {et~al.}(2006){Heymans}, {Van Waerbeke}, {Bacon}, {Berge},
  {Bernstein}, {Bertin}, {Bridle}, {Brown}, {Clowe}, {Dahle}, {Erben}, {Gray},
  {Hetterscheidt}, {Hoekstra}, {Hudelot}, {Jarvis}, {Kuijken}, {Margoniner},
  {Massey}, {Mellier}, {Nakajima}, {Refregier}, {Rhodes}, {Schrabback}, \&
  {Wittman}}]{STEP1}
{Heymans}, C., {Van Waerbeke}, L., {Bacon}, D., {et~al.} 2006, \mnras, 368,
  1323

\bibitem[{{Jain} \& {Taylor}(2003)}]{2003PhRvL..91n1302J}
{Jain}, B. \& {Taylor}, A. 2003, Physical Review Letters, 91, 141302

\bibitem[{{Kaiser} \& {Peacock}(2016)}]{2016MNRAS.455.4518K}
{Kaiser}, N. \& {Peacock}, J.~A. 2016, \mnras, 455, 4518

\bibitem[{{Kilbinger}(2015)}]{Kilbinger-review}
{Kilbinger}, M. 2015, Reports on Progress in Physics, 78, 086901

\bibitem[{{Kitching} {et~al.}(2015){Kitching}, {Viola}, {Hildebrandt}, {Choi},
  {Erben}, {Gilbank}, {Heymans}, {Miller}, {Nakajima}, \& {van
  Uitert}}]{2015arXiv151203627K}
{Kitching}, T.~D., {Viola}, M., {Hildebrandt}, H., {et~al.} 2015, ArXiv
  e-prints

\bibitem[{{Mandelbaum} {et~al.}(2015){Mandelbaum}, {Rowe}, {Armstrong}, {Bard},
  {Bertin}, {Bosch}, {Boutigny}, {Courbin}, {Dawson}, {Donnarumma}, {Fenech
  Conti}, {Gavazzi}, {Gentile}, {Gill}, {Hogg}, {Huff}, {Jee}, {Kacprzak},
  {Kilbinger}, {Kuntzer}, {Lang}, {Luo}, {March}, {Marshall}, {Meyers},
  {Miller}, {Miyatake}, {Nakajima}, {Ngol{\'e} Mboula}, {Nurbaeva}, {Okura},
  {Paulin-Henriksson}, {Rhodes}, {Schneider}, {Shan}, {Sheldon}, {Simet},
  {Starck}, {Sureau}, {Tewes}, {Zarb Adami}, {Zhang}, \& {Zuntz}}]{GREAT3}
{Mandelbaum}, R., {Rowe}, B., {Armstrong}, R., {et~al.} 2015, \mnras, 450, 2963

\bibitem[{{Massey} {et~al.}(2007){Massey}, {Heymans}, {Berg{\'e}}, {Bernstein},
  {Bridle}, {Clowe}, {Dahle}, {Ellis}, {Erben}, {Hetterscheidt}, {High},
  {Hirata}, {Hoekstra}, {Hudelot}, {Jarvis}, {Johnston}, {Kuijken},
  {Margoniner}, {Mandelbaum}, {Mellier}, {Nakajima}, {Paulin-Henriksson},
  {Peeples}, {Roat}, {Refregier}, {Rhodes}, {Schrabback}, {Schirmer}, {Seljak},
  {Semboloni}, \& {van Waerbeke}}]{STEP2}
{Massey}, R., {Heymans}, C., {Berg{\'e}}, J., {et~al.} 2007, \mnras, 376, 13

\bibitem[{{Melchior} \& {Viola}(2012)}]{2012MNRAS.424.2757M}
{Melchior}, P. \& {Viola}, M. 2012, \mnras, 424, 2757

\bibitem[{{Miller} {et~al.}(2013){Miller}, {Heymans}, {Kitching}, {van
  Waerbeke}, {Erben}, {Hildebrandt}, {Hoekstra}, {Mellier}, {Rowe}, {Coupon},
  {Dietrich}, {Fu}, {Harnois-D{\'e}raps}, {Hudson}, {Kilbinger}, {Kuijken},
  {Schrabback}, {Semboloni}, {Vafaei}, \& {Velander}}]{2013MNRAS.429.2858M}
{Miller}, L., {Heymans}, C., {Kitching}, T.~D., {et~al.} 2013, \mnras, 429,
  2858

\bibitem[{{Pillepich} {et~al.}(2012){Pillepich}, {Porciani}, \&
  {Reiprich}}]{2012MNRAS.422...44P}
{Pillepich}, A., {Porciani}, C., \& {Reiprich}, T.~H. 2012, \mnras, 422, 44

\bibitem[{{Rozo} \& {Schmidt}(2010)}]{2010arXiv1009.5735R}
{Rozo}, E. \& {Schmidt}, F. 2010, ArXiv e-prints

\bibitem[{{Schneider} {et~al.}(2015){Schneider}, {Hogg}, {Marshall}, {Dawson},
  {Meyers}, {Bard}, \& {Lang}}]{2015ApJ...807...87S}
{Schneider}, M.~D., {Hogg}, D.~W., {Marshall}, P.~J., {et~al.} 2015, \apj, 807,
  87

\bibitem[{{Schneider} {et~al.}(1992){Schneider}, {Ehlers}, \& {Falco}}]{SEF}
{Schneider}, P., {Ehlers}, J., \& {Falco}, E.~E. 1992, {Gravitational Lenses}

\bibitem[{{Seitz} \& {Schneider}(1997)}]{SeitzS97}
{Seitz}, C. \& {Schneider}, P. 1997, \aap, 318, 687

\bibitem[{{Seitz} {et~al.}(1994){Seitz}, {Schneider}, \& {Ehlers}}]{SSE}
{Seitz}, S., {Schneider}, P., \& {Ehlers}, J. 1994, Classical and Quantum
  Gravity, 11, 2345

\bibitem[{{Taylor} {et~al.}(2007){Taylor}, {Kitching}, {Bacon}, \&
  {Heavens}}]{2007MNRAS.374.1377T}
{Taylor}, A.~N., {Kitching}, T.~D., {Bacon}, D.~J., \& {Heavens}, A.~F. 2007,
  \mnras, 374, 1377

\bibitem[{{Voigt} \& {Bridle}(2010)}]{2010MNRAS.404..458V}
{Voigt}, L.~M. \& {Bridle}, S.~L. 2010, \mnras, 404, 458

\end{thebibliography}



\end{document}